\documentclass[aps,pra,reprint,superscriptaddress]{revtex4-2}

\usepackage{graphicx}
\usepackage{dcolumn}
\usepackage{bm}
\usepackage{amsmath,amssymb}
\usepackage{amsthm}
\usepackage{braket}
\usepackage{url}

\usepackage{hyperref}

\begin{document}

\title{Time of arrival in the semiclassical regime for Gaussian wave packets}

\author{Mathieu Beau}
\affiliation{Physics department, University of Massachusetts, Boston, Massachusetts 02125, USA}

\author{Maximilien Barbier}
\affiliation{Scottish Universities Physics Alliance, University of the West of Scotland, Paisley PA1 2BE, Scotland, United Kingdom}

\author{Vincent Bucourt}
\affiliation{Somerville High School, Somerville, Massachusetts, USA}

\date{\today}

\begin{abstract}
We study the time-of-arrival (TOA) distribution of a Gaussian wave packet in the
semiclassical regime for a general one-dimensional quadratic Hamiltonian, namely a forced
harmonic oscillator with time-dependent frequency. Expanding the Gaussian probability
density as a series in derivatives of the Dirac delta and using the standard rules for the
composition of distributions with the classical trajectory, we obtain compact closed-form
expressions for the leading semiclassical mean arrival time and its standard deviation. The
mean acquires a quantum shift of order $\sigma^2$ set by the classical velocity and
acceleration and by the time derivative of the Ermakov width, while the spread reduces to
the simple form $\Delta\mathcal{T}_x\simeq\sigma(t_x)/v_x$. The latter implies a
time--position uncertainty relation that depends only on the fundamental solutions of the
classical equation of motion. We apply the framework to free fall and to two models of a
time-dependent opening trap under gravity, and we validate every expression against exact
numerical integration of the TOA distribution. These results extend earlier free-fall
predictions to realistic settings in which the trapping potential cannot be neglected.
\end{abstract}

\maketitle

\section{Introduction}

The status of time as an observable in quantum mechanics remains unsettled. Pauli's
argument that no self-adjoint time operator is canonically conjugate to a
lower-bounded Hamiltonian~\cite{pauli1933handbuch} rules out a naive time observable, and
a large body of work has since addressed how a time of arrival (TOA) can nonetheless be
defined and computed~\cite{Allcock69,Muga1,Muga2,MugaLeavens00}, including the axiomatic
construction of a positive-operator-valued measure for the arrival
time~\cite{Kijowski74} and approaches based on detector
models~\cite{AnastopoulosSavvidou12}, decoherent histories~\cite{HalliwellYearsley09}, and
Bohmian trajectories~\cite{das2019arrival,goldstein2024spin}. Among the
available approaches, one route infers
the distribution of a time measurement $\mathcal{T}_x$ at a fixed position $x$ from the
Born-rule distribution of the position measurement $\hat{x}_t$ at fixed time
$t$~\cite{Beau24,Beau24_2,Beau25_2}. This construction is operationally transparent for
Gaussian states and yields explicit predictions that can be confronted with experiment.

A particularly clean application is free fall. For a Gaussian wave packet released in a
uniform gravitational field, the mean arrival time is shifted relative to the classical
value by a mass-dependent quantum correction, and the arrival-time and position
uncertainties obey a reciprocal relation~\cite{Beau24,Beau25_2}. Such effects are within
reach of matter-wave platforms that combine long free-fall times with Bose--Einstein
condensates, including drop-tower experiments~\cite{MicrogravityEarth10}, sounding-rocket
and space missions~\cite{MAIUS18,CAL23}, and proposals to test the universality of free
fall and the equivalence principle~\cite{MicrogravityEarth13,Altschul15,GBAR22}.

The explicit results obtained so far are, however, restricted to free fall, where the
classical trajectory and the wave-packet width are known in closed form. In any real
apparatus the particle is prepared in a trap, and the residual or time-dependent
confinement during the measurement modifies both the classical arrival time and the
spreading of the packet. The purpose of this paper is to remove that restriction. We work
in the semiclassical regime, in which the wave-packet width is small compared with the
arrival distance, and we treat a general quadratic Hamiltonian with a time-dependent
frequency and an external force. The Gaussian density is expanded in derivatives of the
Dirac delta function, and the resulting distribution in the time variable is evaluated through its
moments. This produces closed-form semiclassical expressions for the mean arrival time and
its standard deviation that require only the solution of two ordinary differential
equations, the classical equation of motion and the Ermakov equation for the width.

The paper is organized as follows. Section~\ref{sec:setup} sets up the TOA distribution for
the time-dependent forced harmonic oscillator. Section~\ref{sec:semiclassical} derives the
general semiclassical expansion, the leading expressions for the mean and the standard
deviation of the arrival time, and the associated time--position uncertainty relation.
Section~\ref{sec:applications} applies the framework to free fall and to two models of an
opening trap under gravity, and compares the analytic predictions with exact numerical
integration. Section~\ref{sec:conclusion} concludes. Technical derivations of the distributional identities, the narrow-Gaussian expansion,
and the late-time expansions are collected in the appendices.

\section{Time of arrival distribution for the time-dependent forced harmonic oscillator}
\label{sec:setup}

Consider a nonrelativistic particle of mass $m$ whose position wave function $\psi(x,t)$ obeys the time-dependent Schr\"odinger equation
\[
i\hbar \frac{\partial \psi}{\partial t} = \left( -\frac{\hbar^2}{2m} \frac{\partial^2}{\partial x^2} + V(x,t) \right) \psi,
\]
with
\[
V(x,t) = \frac{1}{2} m \omega(t)^2 x^2 - m\,a(t)\, x\ ,
\]
where $\omega(t)$ is the time-dependent angular frequency and $a(t)$ is a time-dependent
acceleration field (for instance an electric field or gravity). At $t=0$ the wave packet is taken to be the uncorrelated pure Gaussian 
\begin{equation}\label{Eq:Psi0}
    \psi(x, 0) = \dfrac{1}{(2\pi \sigma^2)^{1/4}} e^{-\frac{(x-x_0)^2}{4\sigma^2}} e^{ikx} \, ,
\end{equation}
where $x_0$ is the initial mean position, $\sigma$ is the initial position standard deviation, and $p_0=\hbar k$ is the initial mean momentum. Since both the initial state and the TDSE propagator are Gaussian, the probability density
retains the Gaussian form~\cite{Husimi53}
\begin{equation}\label{Eq:rho}
    \rho(x,t) =  \dfrac{e^{-\frac{(x-x_c(t))^2}{2\sigma(t)^2}}}{\sqrt{2\pi\sigma(t)^2}}\ .
\end{equation}

It was shown in Refs.~\cite{Beau24,Beau24_2} that, for a Gaussian wave packet, the TOA
distribution at position $x$ is
\begin{equation}\label{Eq:TOAdistribution}
    \pi_{x}(t) = \mathcal{N}\ \left|v_c(t)+\dfrac{\dot{\sigma}(t)}{\sigma(t)}(x-x_c(t))\right| \dfrac{e^{-\frac{(x-x_c(t))^2}{2\sigma(t)^2}}}{\sqrt{2\pi\sigma(t)^2}}\ ,
\end{equation}
where $\mathcal{N}$ is the inverse of the time integral of the absolute value of the probability current and therefore normalizes the distribution~\cite{Beau24_2}, $v_c(t)=\dot{x}_c(t)$ is the classical
velocity, and $x_c(t)$ is the classical position solving
\begin{equation}\label{Eq:ClassicalEq}
\ddot{x}=-\omega(t)^2x + a(t)\ ,
\end{equation}
with $x_c(0)=x_0$ and $v_c(0)=v_0\equiv \hbar k/m$.

The standard deviation of the position obeys the Ermakov equation~\cite{Ermakov1880,Pinney50}
\begin{equation}\label{Eq:Ermakov}
    \ddot{\sigma}(t)+\omega(t)^2\sigma(t)=\dfrac{\hbar^2}{4m^2\sigma(t)^3}\ ,
\end{equation}
with $\sigma(0)=\sigma$ and $\dot{\sigma}(0)=0$. Its solution is~\cite{Pinney50,LewisRiesenfeld69}
\begin{equation}\label{Eq:sigma(t)}
\sigma(t) = \sigma\sqrt{G_1(t)^2+\frac{\hbar^2}{4m^2\sigma^4}G_2(t)^2}\ ,
\end{equation}
where $G_1(t)$ and $G_2(t)$ are the fundamental solutions of the homogeneous equation
[Eq.~\eqref{Eq:ClassicalEq} with $a(t)=0$] satisfying $G_1(0)=1$, $\dot{G}_1(0)=0$,
$G_2(0)=0$, $\dot{G}_2(0)=1$.

\section{Semiclassical limit: general formula}
\label{sec:semiclassical}

\subsection{Semiclassical expansion and distributional interpretation}

Near the crossing $t_x$, the semiclassical regime requires the wave-packet width to be small compared with the spatial scale of the motion; in the applications below this condition is $\sigma_x\ll x$, where $\sigma_x\equiv\sigma(t_x)$. Starting from Eq.~\eqref{Eq:rho}, the probability
density admits the distributional expansion (see Appendix~\ref{Appendix:GaussianToDirac})
\begin{equation}\label{Eq:SeriesGaussian}
\rho(x,t)
\;\approx\;
\delta\!\big(x-x_c(t)\big)
+\frac{1}{2}\sigma(t)^2\,\delta''\!\big(x-x_c(t)\big)
+\cdots \, .
\end{equation}
Combining Eq.~\eqref{Eq:SeriesGaussian} with the definition of the TOA distribution in
Eq.~\eqref{Eq:TOAdistribution} gives, at leading semiclassical order,
\begin{equation}\label{Eq:TOAdistribution:sc}
\pi_x(t)
\approx
\mathcal N\,|v(x,t)|
\left[
\delta\!\big(x-x_c(t)\big)
+\frac{1}{2}\sigma(t)^2\,
\delta''\!\big(x-x_c(t)\big)
+\cdots
\right] .
\end{equation}

Equation~\eqref{Eq:TOAdistribution:sc} is understood in the sense of distributions in the
time variable $t$. The TOA distribution is therefore not interpreted pointwise but only
through its action on test functions, that is through the moments of the arrival time,
\begin{equation}\label{Eq:MomentsTOA}
\mathbb{E}\left(\mathcal T_x^{\,n}\right)
=
\int_0^{\infty} t^n\,\pi_x(t)\,dt \, .
\end{equation}
Within this framework, the leading term in Eq.~\eqref{Eq:TOAdistribution:sc} reproduces the
classical time of arrival $t_x$ defined by $x_c(t_x)=x$, while the higher-order terms
generate quantum corrections involving derivatives of the test function evaluated at $t_x$.

\subsection{Explicit calculation}

We now convert the spatial delta distributions into delta distributions in time. Using the
composition rules derived in Appendix~\ref{Appendix:Expressions}, and writing the classical
arrival condition as $x_c(t_i)=x$ with $v_c(t_i)\neq 0$, one finds
\begin{equation}\label{Eq:delta0}
\delta\big(x-x_c(t)\big)
=
\sum_{i=1}^{n}\frac{1}{|v_c(t_i)|}\,\delta(t-t_i),
\end{equation}
\begin{equation}\label{Eq:delta1}
\delta'\!\big(x-x_c(t)\big)
=
-\sum_{i=1}^{n}
\left[
\frac{v_c(t_i)}{|v_c(t_i)|^3}\,\delta'(t-t_i)
+
\frac{a_c(t_i)}{|v_c(t_i)|^{3}}\,\delta(t-t_i)
\right],
\end{equation}
and
\begin{align}\label{Eq:delta2}
\delta''\big(x-x_c(t)\big)
&=
\sum_{i=1}^{n}
\Bigg[
\frac{1}{|v_c(t_i)|^3}\,\delta''(t-t_i)
\nonumber\\
&+
\frac{3a_c(t_i)v_c(t_i)}{|v_c(t_i)|^{5}}\,\delta'(t-t_i)
\nonumber\\
&+
\frac{3a_c(t_i)^2-v_c(t_i)j_c(t_i)}{|v_c(t_i)|^{5}}\,
\delta(t-t_i)
\Bigg] ,
\end{align}
where the roots $t_i$ satisfy $x_c(t_i)=x$, and $v_c'(t_i)=a_c(t_i)$, $a_c'(t_i)=j_c(t_i)$
denote the classical velocity, acceleration, and jerk.

The exact distribution~\eqref{Eq:TOAdistribution} carries an additional contribution
proportional to $\dot\sigma(t)$. In the semiclassical regime
$|x-x_c(t)|\ll |v_c(t)|/|\dot\sigma(t)/\sigma(t)|$, and assuming a single simple crossing $x_c(t_x)=x$ with $v_x>0$, so that the absolute value can be dropped locally, we expand the prefactor to
first order in $(x-x_c)$,
\begin{equation}
\left|\,v_c(t)+[x-x_c(t)]\frac{\dot\sigma(t)}{\sigma(t)}\right|
=
v_c(t)+b(t)\,(x-x_c(t)),
\end{equation}
with $b(t)\equiv \dot\sigma(t)/\sigma(t)$.

Substituting Eqs.~\eqref{Eq:delta0}--\eqref{Eq:delta2} into the semiclassical
expansion~\eqref{Eq:TOAdistribution:sc} and moving each smooth prefactor $F(t)$ through the
delta derivatives by integration by parts, that is using
\begin{align}
    & F(t)\,\delta(t-t_x)=F_x\,\delta(t-t_x),\nonumber\\
    & F(t)\,\delta'(t-t_x)=F_x\,\delta'(t-t_x)-F_x'\,\delta(t-t_x),\nonumber\\
    & F(t)\,\delta''(t-t_x)=F_x\,\delta''(t-t_x)-2F_x'\,\delta'(t-t_x)+F''_x\,\delta(t-t_x),
\end{align}
with $F_x\equiv F(t_x)$, $F'_x\equiv F'(t_x)$, $F''_x\equiv F''(t_x)$. These identities follow directly by evaluating both sides on a test function and applying Leibniz's rule to the derivatives of $F(t)f(t)$ at $t=t_x$. One then obtains the
explicit distributional representation
\begin{widetext}
\begin{align}\label{Eq:Pi:Semiclass:Gen}
\pi_x(t)\approx\mathcal N\Bigg\{
&\delta(t-t_x)
+\frac{\sigma_x^2}{2\,v_x^{2}}\,\delta''(t-t_x)
+\sigma_x^2A_x\,\delta'(t-t_x)
+\sigma_x^2B_x\,\delta(t-t_x)
+\cdots\Bigg\} ,
\end{align}
\end{widetext}
where $v_x\equiv v_c(t_x)$, $\sigma_x\equiv \sigma(t_x)$, and
\begin{align}
A_x\sigma_x^2&\equiv \dfrac{3G_x a_x}{v_x^4}-\dfrac{2G_x'}{v_x^3} +\dfrac{2H_x}{v_x^2} ,\label{Eq:Ax}\\
B_x\sigma_x^2&\equiv \dfrac{G''_x}{v_x}+\dfrac{G_x}{v_x^5}(3a_x^2-v_xj_x)-\dfrac{3a_x G'_x}{v_x^4}+\dfrac{2a_x H_x}{v_x^3}-\dfrac{2H_x}{v_x^2} ,\label{Eq:Bx}
\end{align}
with $G(t) \equiv \tfrac{1}{2}v_c(t)\sigma(t)^2$, $H(t)\equiv\tfrac{1}{2}b(t)\sigma(t)^2$,
and the subscript $x$ again denoting evaluation at $t_x$. Equation~\eqref{Eq:Pi:Semiclass:Gen}
is the explicit distributional form of the semiclassical TOA distribution, suitable for the
direct evaluation of moments.

\begin{figure*}[h!t]
\centering
\includegraphics[width=\textwidth]{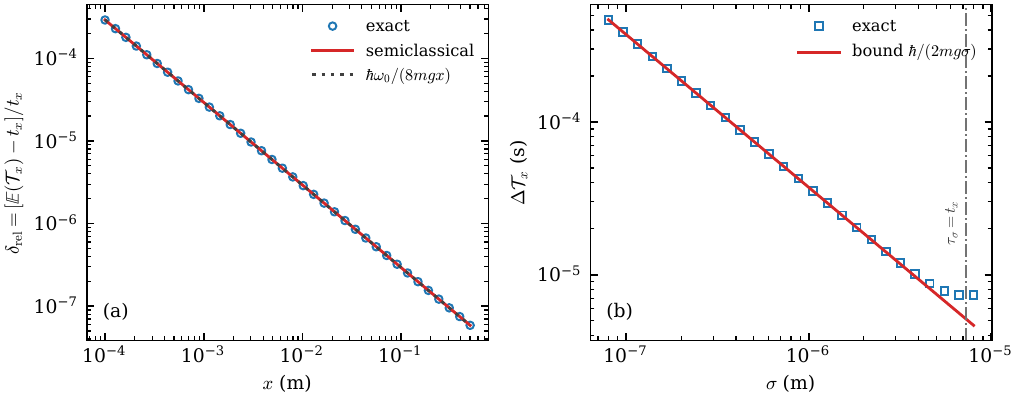}
\caption{\label{Fig:ModelI}
Free fall ($^{87}$Rb, $\omega_0=2\pi\times500\,$Hz).
(a) Relative quantum shift $\delta_{\rm rel}=[\mathbb E(\mathcal T_x)-t_x]/t_x$ versus arrival distance
$x$: exact numerical integration (open circles), semiclassical
formula~\eqref{Eq:FreeFall:Shift:meanTOA} (solid), and the leading harmonic-trap term
$\hbar\omega_0/(8mgx)$ (dotted gray).
(b) Arrival-time spread $\Delta\mathcal T_x$ versus initial width $\sigma$ at $x=0.1\,$m:
exact numerical integration (open squares) and the lower bound $\hbar/(2mg\sigma)$ of
Eq.~\eqref{Eq:TEUR:t0} (solid). The dash-dotted vertical line marks the ballistic onset
$\tau_\sigma=t_x$, beyond which the bound is no longer saturated.}
\end{figure*}

\subsection{Semiclassical mean and standard deviation of the time of arrival}
\label{App:GeneralMoments_corrected}

From Eq.~\eqref{Eq:Pi:Semiclass:Gen} we define the unnormalized moments
$M_k=\int_0^\infty t^k\,\pi_x(t)\,dt$. Using the standard identity
\begin{equation*}
    \int_{0}^{+\infty}\delta^{(p)}(t-t_x)\,f(t)\,dt = (-1)^p f^{(p)}(t_x),
\end{equation*}
for the $p$-th derivative of the Dirac delta, with $t_x>0$, we obtain up to $O(\sigma_x^2)$
\begin{align}
M_0 &\approx 1+\sigma_x^2 B_x,\label{Eq:M0_corrected}\\[0.1cm]
M_1 &\approx t_x+\sigma_x^2\big[B_x t_x-A_x\big],\label{Eq:M1_corrected}\\[0.1cm]
M_2 &\approx t_x^2+\sigma_x^2\big[\tfrac{1}{v_x^2}-2A_x t_x+B_x t_x^2\big],\label{Eq:M2_corrected}
\end{align}
and the normalized moments follow as $\mathbb E(\mathcal T_x^k)=M_k/M_0$.

Expanding $M_1/M_0$ to first order in $\sigma^2$ gives
\begin{equation}
\mathbb E(\mathcal T_x)
\approx
t_x
-\sigma_x^2A_x,
\label{Eq:ET}
\end{equation}
where the $B_x$ contribution cancels against the $\sigma_x^2$ term in the expansion of
$1/M_0\approx 1-\sigma_x^2B_x$. Using the definition~\eqref{Eq:Ax} one finds
$A_x=(a_x-2b_x v_x)/(2v_x^3)$, which yields the compact result
\begin{equation}
\mathbb E(\mathcal T_x)
\approx
t_x
+\sigma_x^2\left(\dfrac{2b_x v_x-a_x}{2v_x^3}\right).
\label{Eq:ET:explicit}
\end{equation}
Thus, by solving the classical equation~\eqref{Eq:ClassicalEq} together with the Ermakov
equation~\eqref{Eq:Ermakov} [equivalently Eq.~\eqref{Eq:sigma(t)}], the leading
$O(\sigma_x^2)$ correction to the mean arrival time follows without evaluating the moment
integral~\eqref{Eq:MomentsTOA} from the exact distribution~\eqref{Eq:TOAdistribution}.
Higher-order corrections can in principle be obtained by retaining higher derivatives in the
asymptotic expansion~\eqref{Eq:SeriesGaussian}.

The second moment is
$\mathbb E(\mathcal T_x^{\,2}) \approx t_x^2 + \sigma_x^2\big(1/v_x^2-2A_x t_x\big)$.
Combining it with Eq.~\eqref{Eq:ET:explicit} and using
$\mathrm{Var}(\mathcal T_x)=\mathbb E(\mathcal T_x^{\,2})-\mathbb E(\mathcal T_x)^2$ with $\Delta\mathcal T_x\equiv\sqrt{\mathrm{Var}(\mathcal T_x)}$, the $A_x$ terms cancel and we
obtain
\begin{equation}
\Delta\mathcal T_x
\approx
\frac{\sigma_x}{v_x} .
\label{Eq:sigmaT}
\end{equation}
This expression is simple and admits a clear classical interpretation: the standard
deviations of the position and of the arrival time are proportional, with the proportionality
coefficient set by the classical velocity at the arrival position $x$. As with the mean, the
result relies only on $\rho(x,t)$ being Gaussian.

Equation~\eqref{Eq:sigmaT} also imposes a lower bound on the product of the arrival-time
uncertainty $\Delta\mathcal T_x$ at position $x$ and the position uncertainty
$\Delta \hat{x}_t$ at time $t$,
\begin{equation}\label{Eq:TEUR:General}
    \Delta\mathcal T_x\,\Delta \hat{x}_t
    \geq
    \dfrac{\hbar}{2\,m\,v_x}
    \left(
        |G_1(t)G_2(t_x)|
        +
        |G_2(t)G_1(t_x)|
    \right).
\end{equation}
The bound follows by minimizing $\Delta\mathcal T_x\,\Delta\hat{x}_t=\sigma(t_x)\sigma(t)/v_x$
over the initial spread $\sigma$ using Eq.~\eqref{Eq:sigma(t)}. The minimizer is
\[
\sigma_{\text{min}}^2
=
\frac{\hbar}{2m}
\sqrt{
\frac{|G_2(t)G_2(t_x)|}{|G_1(t)G_1(t_x)|}
}\ ,
\]
after which Eq.~\eqref{Eq:TEUR:General} follows by direct substitution. In particular, when
the initial position uncertainty is fixed, dropping the $G_1$ term in Eq.~\eqref{Eq:sigma(t)}
gives
\begin{equation}\label{Eq:TEUR:t0}
\Delta\mathcal T_x
\geq
\frac{1}{\sigma}
\frac{\hbar |G_2(t_x)|}{2\,m\,v_x}.
\end{equation}

Time--position uncertainty relations of this kind were obtained in Ref.~\cite{Beau25_2} for
free fall, where the semiclassical and quantum regimes can both be treated explicitly. Here
the result is extended in the semiclassical regime to the class of initially uncorrelated pure Gaussian wave packets evolving in time-dependent harmonic potentials. Taken
together, Eqs.~\eqref{Eq:ET:explicit}--\eqref{Eq:TEUR:t0} generalize the free-fall results of
Refs.~\cite{Beau24,Beau25_2}. Their main interest is that they apply to realistic systems in
which the trapping potential cannot be neglected, as discussed next.

\section{Applications to free fall and a time-dependent harmonic trap under gravity}
\label{sec:applications}

We illustrate the general expressions on three cases: free fall, an opening trap with a
non-resonant frequency, and an opening trap with a resonant frequency. In each case the
semiclassical mean~\eqref{Eq:ET:explicit} and spread~\eqref{Eq:sigmaT} are compared with the
exact result obtained by numerically integrating the moments~\eqref{Eq:MomentsTOA} of the
full distribution~\eqref{Eq:TOAdistribution}. Unless stated otherwise we take $^{87}$Rb,
$m=1.44\times10^{-25}\,$kg; the microgravity estimates of
Sec.~\ref{sec:experimental} use metastable helium, which is routinely employed in cold-atom experiments~\cite{Vassen12}.

\subsection{Model I: free fall}

For free fall $x_c(t)=gt^2/2$ and $v_c(t)=gt$, with $\sigma(t)=\sigma\sqrt{1+t^2/\tau_\sigma^2}$
and $\tau_\sigma=2m\sigma^2/\hbar$. The unique crossing time is $t_x=\sqrt{2x/g}$.
Substituting into Eqs.~\eqref{Eq:ET:explicit} and~\eqref{Eq:sigmaT} gives
\begin{subequations}
\begin{equation}\label{Eq:FreeFall:meanTOA}
     \mathbb E(\mathcal T_x) \approx t_x + \dfrac{\sigma^2}{2g^2\tau_\sigma^2 t_x}\left(1-\dfrac{\tau_\sigma^2}{t_x^2}\right),
\end{equation}
\begin{equation}\label{Eq:FreeFall:stdevTOA}
     \Delta\mathcal T_x \approx \dfrac{\hbar}{2mg\sigma} .
\end{equation}
\end{subequations}
The relative quantum shift of the mean is therefore
\begin{equation}\label{Eq:FreeFall:Shift:meanTOA}
\delta_{\rm rel} \equiv \frac{\mathbb E(\mathcal T_x)-t_x}{t_x}
=\frac{\hbar^2}{16gxm^2\sigma^2}-\frac{\sigma^2}{8x^2}.
\end{equation}
In the long-time regime $t_x\gg \tau_\sigma$ and the far-field regime $x\gg \sigma$, the
second term is negligible, so the arrival is delayed relative to the classical time, in
agreement with Eq.~(18) of Ref.~\cite{Beau24}. Equation~\eqref{Eq:FreeFall:stdevTOA}
saturates the bound~\eqref{Eq:TEUR:t0} with $G_2(t_x)=t_x$ and $v_x=gt_x$: the initial
spread of the wave function sets the minimum measurable arrival-time uncertainty.

Figure~\ref{Fig:ModelI} confirms both predictions against exact numerical integration. The
relative shift $\delta_{\rm rel}(x)$ in panel (a) follows the $1/x$ scaling of the leading term over
the entire range, with the semiclassical curve tracking the exact result to a relative
accuracy below $10^{-6}$. Panel (b) shows the arrival-time spread as a function of the
initial width $\sigma$ at fixed $x$. The exact result follows the bound $\hbar/(2mg\sigma)$
as long as the packet reaches the ballistic regime $\tau_\sigma\ll t_x$, and departs from it
once $\tau_\sigma\gtrsim t_x$, where the packet has not yet entered free spreading at the
detector.

When the packet is prepared in the trap, its initial width is $\sigma=\sqrt{\hbar/(2m\omega_0)}$,
and Eq.~\eqref{Eq:FreeFall:Shift:meanTOA} becomes
\begin{equation}\label{Eq:FreeFall:Shift:meanTOA:HO}
\delta_{\rm rel}=\dfrac{\hbar\,\omega_0}{8\,mgx}
\left(1- \dfrac{g}{2x\,\omega_0^2} \right).
\end{equation}
The prefactor is the ratio between the ground-state energy $\hbar\omega_0/2$ and four times
the gravitational potential energy $mgx$, so the dynamics is semiclassical when the quantum
energy scale is small compared with the gravitational one. The second term,
$g/(2x\omega_0^2)=mgx/(2m\omega_0^2x^2)$, is purely classical and compares the gravitational
energy with four times the harmonic energy of the initial trap at $x$. For typical Earth
settings, $\omega_0\sim10^4$--$10^5\,$Hz and $g\sim10\,$m\,s$^{-2}$, so as long as
$x\gg10^{-8}\,$m the second term is negligible and
$\delta_{\rm rel}\approx \hbar\omega_0/(8mgx)$ gives the relative quantum correction.

Reaching a regime where quantum effects are sizable requires
$\hbar\omega_0\sim mgx$, which can be approached by reducing the free-fall height or the
particle mass, lowering the effective gravity in microgravity or drop-tower
settings~\cite{MicrogravityEarth10,MAIUS18,CAL23}, or increasing the trap frequency
$\omega_0$. Finally, the ratio of the spread to the classical time is
\[
    \dfrac{\Delta\mathcal{T}_x}{t_x}\approx \sqrt{2\delta_{\rm rel}}= \sqrt{\dfrac{\hbar\omega_0}{4mgx}},
\]
so the relative arrival-time uncertainty is again governed by the ratio between the trap
ground-state energy and the gravitational energy.

\begin{figure}[ht]
\centering
\includegraphics[width=\columnwidth]{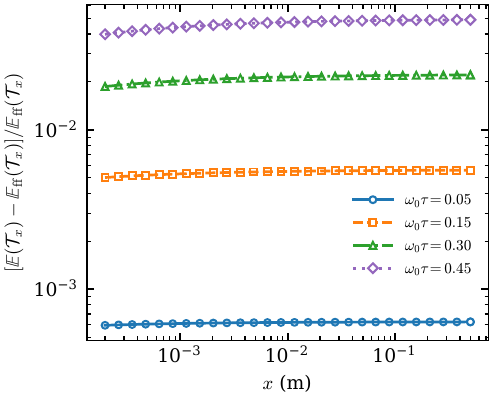}
\caption{\label{Fig:ModelII}
Model II, convergence to free fall as $\tau\to0$. Relative deviation of the mean arrival
time from the free-fall value,
$[\mathbb E(\mathcal T_x)-\mathbb E_{\rm ff}(\mathcal T_x)]/\mathbb E_{\rm ff}(\mathcal T_x)$,
versus arrival distance $x$, for $\omega_0\tau=0.05,0.15,0.30,0.45$
($\omega_0=2\pi\times500\,$Hz). Open markers denote exact numerical integration and the
corresponding lines denote the semiclassical formula~\eqref{Eq:ET:explicit}; marker shapes and
line styles identify the four values of $\omega_0\tau$ as shown in the legend. The deviation
scales as $(\omega_0\tau)^2$, consistent with
$g_{\rm eff}=g(1-\tfrac12\omega_0^2\tau^2)$.}
\end{figure}

\begin{figure}[ht]
\centering
\includegraphics[width=\columnwidth]{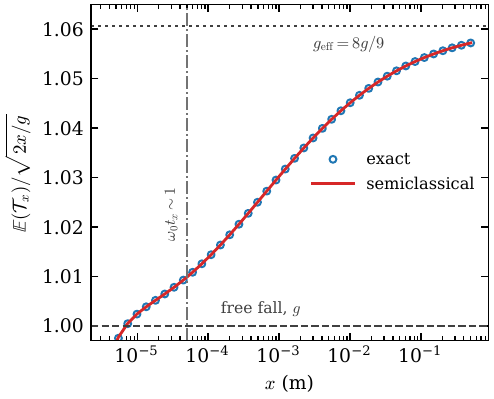}
\caption{\label{Fig:ModelIII}
Model III, crossover from free fall to effective gravity. Ratio of the mean arrival time to
the free-fall value $\sqrt{2x/g}$ versus arrival distance $x$, for
$\omega_0=2\pi\times50\,$Hz. Open circles denote exact numerical integration and the solid
line denotes the semiclassical formula~\eqref{Eq:ET:explicit}. Dashed and dotted horizontal
lines mark the free-fall ($g$) and effective-gravity
($g_{\rm eff}=\tfrac{8}{9}g$, ratio $\sqrt{9/8}$) limits; the dash-dotted vertical line marks
$\omega_0t_x\sim1$.}
\end{figure}

\subsection{Model II: non-resonant opening trap under gravity}

We next take
\[
    \omega(t) = \omega_0\dfrac{\tau}{t+\tau}\ ,
\]
with $\tau < 1/(2\omega_0)$, which keeps the motion overdamped so that the
crossing time $t_x$ defined by $x_c(t_x)=x$ is unique. Time-dependent frequency protocols
of this type are routinely engineered in trap-decompression and shortcut-to-adiabaticity
schemes~\cite{Chen10traps,GueryOdelin19}. We solve Eqs.~\eqref{Eq:ClassicalEq}
and~\eqref{Eq:Ermakov} numerically for $t_x,v_x,a_x,\sigma_x,b_x$ and substitute into
Eqs.~\eqref{Eq:ET:explicit} and~\eqref{Eq:sigmaT}.

In the fast-opening regime $\tau\ll t_x$ and $\omega_0\tau\ll 1$, the classical trajectory
reduces to an effective free fall (see Appendix~\ref{Appendix:LateTime}),
\begin{equation}
x_c(t)\simeq \frac{g_{\rm eff}}{2}\,t^2,\qquad v_c(t)\simeq g_{\rm eff}\,t,
\qquad
g_{\rm eff}=g\Big(1-\tfrac12\omega_0^2\tau^2\Big),
\label{eq:geff_ff}
\end{equation}
while the width approaches the free-particle form
\begin{equation}
\sigma(t)\simeq \sigma_{\rm free}(t)=\sigma\sqrt{1+\frac{\hbar^2 t^2}{4m^2\sigma^4}},
\label{eq:sigma_free_ff}
\end{equation}
up to relative corrections of order $(\omega_0\tau)^2\ln(t/\tau)$. Thus to leading
order in $(\omega_0\tau)^2$ the semiclassical mean and spread coincide with the free-fall
expressions~\eqref{Eq:FreeFall:meanTOA} and~\eqref{Eq:FreeFall:stdevTOA} under the
replacements $g\to g_{\rm eff}$ and $\sigma(t)\to\sigma_{\rm free}(t)$.

Figure~\ref{Fig:ModelII} confirms this picture. The relative deviation of the mean arrival
time from the free-fall value is nearly independent of $x$ and decreases as $(\omega_0\tau)^2$,
from about $5\times10^{-2}$ at $\omega_0\tau=0.45$ to about $6\times10^{-4}$ at
$\omega_0\tau=0.05$, in quantitative agreement with the effective-gravity shift
$\tfrac14(\omega_0\tau)^2$ predicted by Eq.~\eqref{eq:geff_ff}. The semiclassical and exact
results are indistinguishable on the scale of the figure.

\begin{figure*}[h!t]
\centering
\includegraphics[width=\textwidth]{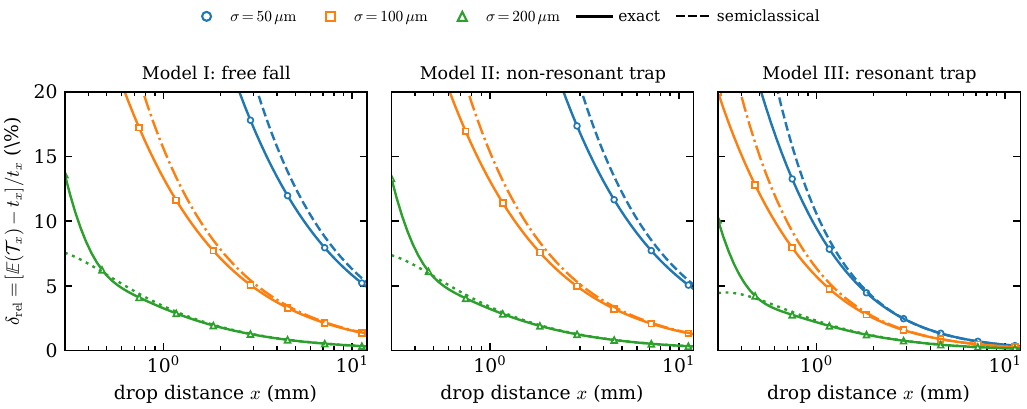}
\caption{\label{Fig:models}
Relative mean time-of-arrival shift
$\delta_{\rm rel}=[\mathbb E(\mathcal T_x)-t_x]/t_x$ versus drop distance $x$ (logarithmic
scale) for the three models, for a metastable-helium packet
($^4$He$^*$, $m=6.64\times10^{-27}\,$kg) in low gravity $g=10^{-5}\,\mathrm{m\,s^{-2}}$,
prepared at initial widths $\sigma=50,100,200\,\mu$m (trap frequencies
$\omega_0/2\pi=0.51,0.13,0.032\,$Hz, from $\sigma=\sqrt{\hbar/(2m\omega_0)}$). Model~II uses
$\omega_0\tau=0.05$. Solid curves with open circle, square, and triangle markers denote
exact integration of the arrival-time distribution~\eqref{Eq:TOAdistribution}; dashed curves denote the semiclassical formula~\eqref{Eq:ET:explicit}. The global legend identifies
both the initial widths and the two methods. Models~I and~II nearly coincide and reach the
percent level at millimeter drops; the resonant trap (Model~III) suppresses the shift through
$g_{\rm eff}=\tfrac{8}{9}g$ and the reshaping of $\sigma(t)$. Exact and semiclassical curves
agree where $\sigma_x\ll x$ (large $x$) and separate as $\sigma_x\to x$ (small $x$).}
\end{figure*}

\subsection{Model III: resonant opening trap under gravity}

We finally consider Eq.~\eqref{Eq:ClassicalEq} with
\[
\omega(t) = \dfrac{\omega_0}{1+2\omega_0 t}\ , \qquad a(t)=g,
\]
and $x(0)=x_0$, $\dot x(0)=v_0$. The two homogeneous solutions with $G_1(0)=1,\dot G_1(0)=0$
and $G_2(0)=0,\dot G_2(0)=1$ are
\begin{subequations}\label{Eq:ModelIII:Gs}
\begin{equation}\label{Eq:ModelIII:G1}
    G_1(t) = \tfrac{1}{2} \sqrt{1 + 2 \omega_0 t}\, \big[2 - \ln(1 + 2 \omega_0 t)\big],
\end{equation}
\begin{equation}\label{Eq:ModelIII:G2}
G_2(t) = \frac{\sqrt{1 + 2 \omega_0 t}\,\ln(1 + 2 \omega_0 t)}{2\omega_0}.
\end{equation}
\end{subequations}
The full solution is
\begin{align}\label{Eq:ModelIII:Xc}
x_c(t) &= \left(x_0-\dfrac{g}{9\omega_0^2}\right)G_1(t)+\left(v_0-\dfrac{4g}{9\omega_0}\right)G_2(t) \nonumber \\
&+\dfrac{g}{9\omega_0^2}(1+2\omega_0 t)^2\ ,
\end{align}
the last term being a particular solution. The crossing time $t_x$ has no closed form but is
readily obtained numerically; $v_x$ and $a_x$ then follow from
Eqs.~\eqref{Eq:ModelIII:Gs}--\eqref{Eq:ModelIII:Xc}, and $\sigma_x,b_x$ from
Eq.~\eqref{Eq:sigma(t)}. Substitution into Eqs.~\eqref{Eq:ET:explicit}
and~\eqref{Eq:sigmaT} gives the semiclassical mean and spread.

Two regimes emerge. For $\omega_0 t_x\ll 1$ the trajectory and width approach the free-fall
forms, because the trap length $\sqrt{\hbar/(2m\omega_0)}$ is large compared with the
ballistic spreading length $\lambda_x\equiv\sqrt{\hbar t_x/m}$, so the harmonic potential
acts weakly during the fall; the corrections to $x_c(t)$ and $\sigma(t)$ are of order
$(\omega_0 t_x)^2$. For $\omega_0 t_x\gg 1$ the trap length is small compared with
$\lambda_x$ and the confinement is significant. Keeping the leading term of
Eq.~\eqref{Eq:ModelIII:Xc}, $x_c(t)\approx \tfrac12 g_{\rm eff}t^2$ with
$g_{\rm eff}=\tfrac{8}{9}g$, gives the closed-form estimate $t_x=\sqrt{2x/g_{\rm eff}}$,
which exceeds the free-fall time $\sqrt{2x/g}$ because the trap slows the fall.

Figure~\ref{Fig:ModelIII} shows the crossover between these regimes through the ratio
$\mathbb E(\mathcal T_x)/\sqrt{2x/g}$, which interpolates between unity (free fall, small $x$)
and $\sqrt{9/8}\simeq1.061$ (effective gravity $g_{\rm eff}=\tfrac{8}{9}g$, large $x$), with
the transition near $\omega_0 t_x\sim1$. The semiclassical curve reproduces the exact result
throughout, including the crossover.

\subsection{Experimental estimates in microgravity}
\label{sec:experimental}

The relative shift $\delta_{\rm rel}$ becomes sizable when the trap ground-state energy
$\hbar\omega_0/2$ is not negligible compared with the gravitational energy $mgx$. As
noted above, this is favored by a light mass, a low effective gravity, and a loose
trap. A realistic setting that combines all three is metastable helium
($^4$He$^*$, $m=6.64\times10^{-27}\,$kg) released from a decompressed trap on a
low-gravity platform, such as a drop tower, a sounding rocket, or a space
station~\cite{MicrogravityEarth10,MAIUS18,CAL23}, where single-atom detection of $^4$He$^*$
is available~\cite{Vassen12} and time-resolved arrival distributions are measured routinely
on microchannel-plate detectors~\cite{Schellekens05}. Initial widths in the range considered
below are attainable by delta-kick collimation and matter-wave
lensing~\cite{MicrogravityEarth13,Kovachy15,CAL22}.

Figure~\ref{Fig:models} compares the three models in this regime, taking
$g=10^{-5}\,\mathrm{m\,s^{-2}}$ and initial widths $\sigma=50,100,200\,\mu$m, which
correspond through $\sigma=\sqrt{\hbar/(2m\omega_0)}$ to trap frequencies
$\omega_0/2\pi=0.51,0.13,0.032\,$Hz. Free fall (Model~I) and the non-resonant opening trap
(Model~II, evaluated at $\omega_0\tau=0.05$) are nearly indistinguishable and reach the
percent level at millimeter drops; for $\sigma=50\,\mu$m the shift is about $9\%$ at
$x=6\,$mm. The resonant trap (Model~III) suppresses the shift by up to an order of
magnitude, through the slower effective fall $g_{\rm eff}=\tfrac{8}{9}g$ and the reshaping of
$\sigma(t)$ by the confinement. In every case the semiclassical
prediction~\eqref{Eq:ET:explicit} tracks the exact integration of
Eq.~\eqref{Eq:TOAdistribution} at large $x$, where $\sigma_x\ll x$, and departs from it at
small $x$, where the packet becomes as wide as the drop distance and the semiclassical
expansion breaks down. The corresponding free-fall times $t_x=\sqrt{2x/g}$ range from about
$14\,$s at $x=1\,$mm to $45\,$s at $x=10\,$mm, within reach of space-based platforms.

\section{Conclusion}
\label{sec:conclusion}

We have derived compact semiclassical expressions for the mean and the standard deviation of
the quantum time of arrival of a Gaussian wave packet evolving under a general
one-dimensional quadratic Hamiltonian. The mean acquires an $O(\sigma^2)$ quantum shift set
by the classical velocity and acceleration and by the Ermakov width, while the spread takes
the form $\Delta\mathcal T_x\simeq\sigma(t_x)/v_x$, which yields a time--position uncertainty
relation governed by the fundamental solutions of the classical equation of motion. The
expressions require only the solution of the classical equation of motion and the Ermakov
equation, and they reduce to the known free-fall results in the appropriate limit. Applied
to two models of a time-dependent opening trap under gravity, they reproduce exact numerical
integration across all regimes, including the convergence of the non-resonant model to free
fall and the crossover of the resonant model to an effective gravity $g_{\rm eff}=\tfrac{8}{9}g$.
Because the results hold for the class of initially uncorrelated pure Gaussian wave packets
considered here in a quadratic potential, they provide arrival-time predictions for realistic
preparations in which the trap cannot be neglected,
which is the situation of interest for free-fall and equivalence-principle tests with
matter waves.

\bibliography{Ref}

@book{pauli1933handbuch,
  author    = {Pauli, Wolfgang and others},
  title     = {Handbuch der Physik},
  publisher = {Geiger and Scheel},
  volume    = {2},
  pages     = {83},
  year      = {1933}
}

@article{Allcock69,
  author  = {Allcock, G. R.},
  title   = {The time of arrival in quantum mechanics I. Formal considerations},
  journal = {Annals of Physics},
  volume  = {53},
  pages   = {253--285},
  year    = {1969},
  note    = {See also ibid. 53, 286 (1969); 53, 311 (1969)}
}

@book{Muga1,
  editor    = {Muga, J. G. and Mayato, R. S. and Egusquiza, I. L.},
  title     = {Time in Quantum Mechanics -- Vol. 1},
  edition   = {2},
  series    = {Lecture Notes in Physics},
  volume    = {734},
  publisher = {Springer},
  address   = {Berlin, Heidelberg},
  year      = {2008}
}

@book{Muga2,
  editor    = {Muga, G. and Ruschhaupt, A. and del Campo, A.},
  title     = {Time in Quantum Mechanics -- Vol. 2},
  edition   = {1},
  series    = {Lecture Notes in Physics},
  volume    = {789},
  publisher = {Springer},
  address   = {Berlin, Heidelberg},
  year      = {2009}
}

@article{MugaLeavens00,
  author  = {Muga, J. G. and Leavens, C. R.},
  title   = {Arrival time in quantum mechanics},
  journal = {Physics Reports},
  volume  = {338},
  pages   = {353--438},
  year    = {2000}
}

@article{Kijowski74,
  author  = {Kijowski, J.},
  title   = {On the time operator in quantum mechanics and the Heisenberg uncertainty relation for energy and time},
  journal = {Reports on Mathematical Physics},
  volume  = {6},
  pages   = {361--386},
  year    = {1974}
}

@article{AnastopoulosSavvidou12,
  author  = {Anastopoulos, C. and Savvidou, N.},
  title   = {Time-of-arrival probabilities for general particle detectors},
  journal = {Physical Review A},
  volume  = {86},
  pages   = {012111},
  year    = {2012}
}

@article{HalliwellYearsley09,
  author  = {Halliwell, J. J. and Yearsley, J. M.},
  title   = {Arrival times, complex potentials, and decoherent histories},
  journal = {Physical Review A},
  volume  = {79},
  pages   = {062101},
  year    = {2009}
}

@article{das2019arrival,
  author  = {Das, S. and D{\"u}rr, D.},
  title   = {Arrival time distributions of spin-1/2 particles},
  journal = {Scientific Reports},
  volume  = {9},
  pages   = {2242},
  year    = {2019}
}

@article{goldstein2024spin,
  author  = {Goldstein, S. and Tumulka, R. and Zangh{\`i}, N.},
  title   = {On the spin dependence of detection times and the nonmeasurability of arrival times},
  journal = {Scientific Reports},
  volume  = {14},
  pages   = {3775},
  year    = {2024}
}

@article{Beau24,
  author  = {Beau, M. and Martellini, L.},
  title   = {Quantum delay in the time of arrival of free-falling atoms},
  journal = {Physical Review A},
  volume  = {109},
  pages   = {012216},
  year    = {2024}
}

@article{Beau24_2,
  author  = {Beau, M. and Barbier, M. and Martellini, R. and Martellini, L.},
  title   = {Time-of-arrival distributions for continuous quantum systems and application to quantum backflow},
  journal = {Physical Review A},
  volume  = {110},
  pages   = {052217},
  year    = {2024}
}

@article{Beau25_2,
  author  = {Beau, M. and Szczepanski, T. and Martellini, R. and Martellini, L.},
  title   = {Quantum arrival times in free fall},
  journal = {Journal of Physics: Conference Series},
  volume  = {3017},
  pages   = {012022},
  year    = {2025}
}

@article{MicrogravityEarth10,
  author  = {van Zoest, T. and others},
  title   = {Bose-Einstein condensation in microgravity},
  journal = {Science},
  volume  = {328},
  pages   = {1540--1543},
  year    = {2010}
}

@article{MAIUS18,
  author  = {Becker, D. and others},
  title   = {Space-borne Bose--Einstein condensation for precision interferometry},
  journal = {Nature},
  volume  = {562},
  pages   = {391--395},
  year    = {2018}
}

@article{CAL23,
  author  = {Elliott, E. R. and others},
  title   = {Quantum gas mixtures and dual-species atom interferometry in space},
  journal = {Nature},
  volume  = {623},
  pages   = {502--508},
  year    = {2023}
}

@article{MicrogravityEarth13,
  author  = {M{\"u}ntinga, H. and others},
  title   = {Interferometry with Bose--Einstein condensates in microgravity},
  journal = {Physical Review Letters},
  volume  = {110},
  pages   = {093602},
  year    = {2013}
}

@article{Altschul15,
  author  = {Altschul, B. and others},
  title   = {Quantum tests of the Einstein equivalence principle with the STE--QUEST space mission},
  journal = {Advances in Space Research},
  volume  = {55},
  pages   = {501--524},
  year    = {2015}
}

@article{GBAR22,
  author  = {Rousselle, O. and Clad{\'e}, P. and Guellati-Khelifa, S. and Gu{\'e}rout, R. and Reynaud, S.},
  title   = {Analysis of the timing of freely falling antihydrogen},
  journal = {New Journal of Physics},
  volume  = {24},
  pages   = {033045},
  year    = {2022}
}

@article{Husimi53,
  author  = {Husimi, K.},
  title   = {Miscellanea in elementary quantum mechanics, II},
  journal = {Progress of Theoretical Physics},
  volume  = {9},
  pages   = {381--402},
  year    = {1953}
}

@article{Ermakov1880,
  author  = {Ermakov, V. P.},
  title   = {Second-order differential equations: Conditions of complete integrability},
  journal = {Applicable Analysis and Discrete Mathematics},
  volume  = {2},
  pages   = {123--145},
  year    = {2008},
  note    = {Translation of Univ. Izv. Kiev 20, 1 (1880)}
}

@article{Pinney50,
  author  = {Pinney, E.},
  title   = {The nonlinear differential equation $y''+p(x)y+cy^{-3}=0$},
  journal = {Proceedings of the American Mathematical Society},
  volume  = {1},
  pages   = {681},
  year    = {1950}
}

@article{LewisRiesenfeld69,
  author  = {Lewis, H. R. and Riesenfeld, W. B.},
  title   = {An exact quantum theory of the time-dependent harmonic oscillator and of a charged particle in a time-dependent electromagnetic field},
  journal = {Journal of Mathematical Physics},
  volume  = {10},
  pages   = {1458--1473},
  year    = {1969}
}

@article{Vassen12,
  author  = {Vassen, W. and Cohen-Tannoudji, C. and Leduc, M. and Boiron, D. and Westbrook, C. I. and Truscott, A. and Baldwin, K. and Birkl, G. and Cancio, P. and Trippenbach, M.},
  title   = {Cold and trapped metastable noble gases},
  journal = {Reviews of Modern Physics},
  volume  = {84},
  pages   = {175--210},
  year    = {2012}
}

@article{Chen10traps,
  author  = {Chen, X. and Ruschhaupt, A. and Schmidt, S. and del Campo, A. and Gu{\'e}ry-Odelin, D. and Muga, J. G.},
  title   = {Fast optimal frictionless atom cooling in harmonic traps: Shortcut to adiabaticity},
  journal = {Physical Review Letters},
  volume  = {104},
  pages   = {063002},
  year    = {2010}
}

@article{GueryOdelin19,
  author  = {Gu{\'e}ry-Odelin, D. and Ruschhaupt, A. and Kiely, A. and Torrontegui, E. and Mart{\'i}nez-Garaot, S. and Muga, J. G.},
  title   = {Shortcuts to adiabaticity: Concepts, methods, and applications},
  journal = {Reviews of Modern Physics},
  volume  = {91},
  pages   = {045001},
  year    = {2019}
}

@article{Schellekens05,
  author  = {Schellekens, M. and Hoppeler, R. and Perrin, A. and Viana Gomes, J. and Boiron, D. and Aspect, A. and Westbrook, C. I.},
  title   = {Hanbury Brown Twiss effect for ultracold quantum gases},
  journal = {Science},
  volume  = {310},
  pages   = {648--651},
  year    = {2005}
}

@article{Kovachy15,
  author  = {Kovachy, T. and Hogan, J. M. and Sugarbaker, A. and Dickerson, S. M. and Donnelly, C. A. and Overstreet, C. and Kasevich, M. A.},
  title   = {Matter wave lensing to picokelvin temperatures},
  journal = {Physical Review Letters},
  volume  = {114},
  pages   = {143004},
  year    = {2015}
}

@article{CAL22,
  author  = {Gaaloul, N. and others},
  title   = {A space-based quantum gas laboratory at picokelvin energy scales},
  journal = {Nature Communications},
  volume  = {13},
  pages   = {7889},
  year    = {2022}
}

\newpage
\onecolumngrid

\vspace{5mm} 

\begin{center}
\textbf{\large APPENDIX}
\end{center}

\appendix


\section{Asymptotic expansion of a narrow Gaussian in delta derivatives}
\label{Appendix:GaussianToDirac}

Let
\begin{equation}
 g_\sigma(x)=\frac{1}{\sqrt{2\pi}\,\sigma}\exp\!\left(-\frac{x^2}{2\sigma^2}\right)
\end{equation}
be the normalized centered Gaussian. For every integer $N\geq0$ and every Schwartz test
function $\phi\in\mathcal S(\mathbb R)$,
\begin{equation}
\int_{-\infty}^{\infty}g_\sigma(x)\phi(x)\,dx
=
\sum_{n=0}^{N}\frac{\sigma^{2n}}{2^n n!}\,\phi^{(2n)}(0)
+O(\sigma^{2N+2})
\qquad (\sigma\to0).
\label{eq:GaussianAsymptoticAction}
\end{equation}
Equivalently, in the sense of asymptotic expansions of distributions,
\begin{equation}
 g_\sigma
\sim
\sum_{n=0}^{\infty}\frac{\sigma^{2n}}{2^n n!}\,\delta^{(2n)}
\qquad (\sigma\to0).
\label{eq:GaussianAsymptoticDistribution}
\end{equation}

To prove Eq.~\eqref{eq:GaussianAsymptoticAction}, set $x=\sigma y$ and write
\begin{equation}
\int g_\sigma(x)\phi(x)\,dx
=
\int_{-\infty}^{\infty}\frac{e^{-y^2/2}}{\sqrt{2\pi}}\,\phi(\sigma y)\,dy.
\end{equation}
Taylor's theorem with remainder gives
\begin{equation}
\phi(\sigma y)
=
\sum_{k=0}^{2N+1}\frac{\phi^{(k)}(0)}{k!}(\sigma y)^k
+R_{2N+2}(\sigma y).
\end{equation}
The odd terms vanish after integration, while the Gaussian moments satisfy
\begin{equation}
\int_{-\infty}^{\infty}\frac{e^{-y^2/2}}{\sqrt{2\pi}}y^{2n}\,dy
=\frac{(2n)!}{2^n n!}.
\end{equation}
Because $\phi$ and its derivatives are rapidly decreasing, the integrated Taylor remainder
is $O(\sigma^{2N+2})$. This proves Eq.~\eqref{eq:GaussianAsymptoticAction}; taking $N=1$
yields Eq.~\eqref{Eq:SeriesGaussian} in the main text.

\section{Composition of the Dirac delta with the classical trajectory}
\label{Appendix:Expressions}

Let $u(t)$ be a smooth function with isolated simple zeros $t_i$, so that $u(t_i)=0$ and
$u'(t_i)\neq 0$. The following distributional identities hold:
\begin{equation}
\delta\!\big(u(t)\big)
=
\sum_{i=1}^{n}\frac{1}{|u'(t_i)|}\,\delta(t-t_i),
\label{delta_u_t_standard}
\end{equation}
\begin{equation}
\delta'\!\big(u(t)\big)
=
\sum_{i=1}^{n}\frac{1}{|u'(t_i)|^{3}}
\Big[
u'(t_i)\,\delta'(t-t_i)
+
u''(t_i)\,\delta(t-t_i)
\Big],
\label{first_deriv_delta}
\end{equation}
and
\begin{align}
&\delta''\!\big(u(t)\big)
=
\sum_{i=1}^{n}\frac{1}{|u'(t_i)|^{5}}\times\nonumber\\
&
\Big\{
u'(t_i)\Big[
u'(t_i)\,\delta''(t-t_i)
+3u''(t_i)\,\delta'(t-t_i)
-u'''(t_i)\,\delta(t-t_i)
\Big]
\nonumber\\
&\quad
+3\big[u''(t_i)\big]^2\,\delta(t-t_i)
\Big\}.
\label{sec_deriv_delta}
\end{align}
While the expression~\eqref{delta_u_t_standard} is standard, the proofs of Eqs.~\eqref{first_deriv_delta} and~\eqref{sec_deriv_delta} are given in
Secs.~\ref{proof_deriv_delta} and~\ref{proof_sec_deriv_delta}, respectively. Applying the identities~\eqref{delta_u_t_standard}-\eqref{sec_deriv_delta} to $u(t)=x-x_c(t)$, whose roots $t_i$ satisfy $x_c(t_i)=x$ and
$u'(t_i)=-v_c(t_i)\neq 0$, yields
\begin{equation}\label{Eq:delta0app}
\delta\big(x-x_c(t)\big)
=
\sum_{i=1}^{n}\frac{1}{|v_c(t_i)|}\,\delta(t-t_i),
\end{equation}
\begin{equation}\label{Eq:delta1app}
\delta'\!\big(x-x_c(t)\big)
=
-\sum_{i=1}^{n}
\left[
\frac{v_c(t_i)}{|v_c(t_i)|^3}\,\delta'(t-t_i)
+
\frac{a_c(t_i)}{|v_c(t_i)|^{3}}\,\delta(t-t_i)
\right],
\end{equation}
and
\begin{align}\label{Eq:delta2app}
\delta''\big(x-x_c(t)\big)
=
\sum_{i=1}^{n}
\Bigg[
&\frac{1}{|v_c(t_i)|^3}\,\delta''(t-t_i)
+
\frac{3a_c(t_i)v_c(t_i)}{|v_c(t_i)|^{5}}\,\delta'(t-t_i)
\nonumber\\
&+
\frac{3a_c(t_i)^2-v_c(t_i)j_c(t_i)}{|v_c(t_i)|^{5}}\,
\delta(t-t_i)
\Bigg],
\end{align}
which are Eqs.~\eqref{Eq:delta0}--\eqref{Eq:delta2} of the main text.

\subsection{Proof of the expression of \texorpdfstring{$\delta'[u(t)]$}{the first derivative}}
\label{proof_deriv_delta}

Assume $u(t)$ has a unique zero $t_0$, with
\begin{equation}
    u(t_0) = 0 ,
    \label{t_0_def}
\end{equation}
and
\begin{equation}
    u'(t_0) \neq 0 .
    \label{deriv_u_t_0_nonzero}
\end{equation}
Take an interval $I \subset \mathbb{R}$ with $t_0 \in I$, so that $u : I \to J$ with
$0 \in J$. For an arbitrary test function $f(t)$ on $I$, consider
\begin{equation}
    \mathcal{I} \equiv \int_I dt \, f(t)\, \delta' [u(t)] .
    \label{I_int_def}
\end{equation}
Changing variables to $y = u(t)$, and noting that the relevant neighborhood of $t_0$ can be
taken small enough that $u$ is strictly monotonic [by~\eqref{deriv_u_t_0_nonzero}] and hence
invertible, $t = u^{-1}(y)$, we obtain
\begin{equation*}
    \mathcal{I} = \int_J \frac{dy}{\lvert u'[u^{-1}(y)] \rvert} \, f[u^{-1}(y)]\, \delta' (y) ,
\end{equation*}
that is
\begin{equation}
    \mathcal{I} = \int_J dy \, \phi(y)\, \delta' (y) ,
    \label{I_phi_expr}
\end{equation}
where
\begin{equation}
    \phi(y) \equiv \frac{\phi_1(y)}{\phi_2(y)} ,
    \label{phi_def}
\end{equation}
with
\begin{equation}
    \phi_1(y) \equiv f[u^{-1}(y)] \qquad \text{and} \qquad \phi_2(y) \equiv \lvert u'[u^{-1}(y)] \rvert .
    \label{phi_1_phi_2_def}
\end{equation}
By definition of $\delta'(y)$ and since $0 \in J$,
\begin{equation}
    \mathcal{I} = - \phi'(0) = - \left. \frac{d}{dy} \phi(y) \right\rvert_{y = 0} .
    \label{I_deriv_phi_expr}
\end{equation}
We compute $\phi'(y)$, distinguishing the cases $u'[u^{-1}(y)] > 0$ and $u'[u^{-1}(y)] < 0$,
and use
\begin{equation}
    \frac{d}{dy} u^{-1}(y) = \frac{1}{u'[u^{-1}(y)]} .
    \label{deriv_inverse_fct}
\end{equation}

\subsubsection{Case \texorpdfstring{$u'[u^{-1}(y)] > 0$}{1}}

Here $\phi_2(y) = u'[u^{-1}(y)]$, and in particular $u'(t_0)>0$. From~\eqref{phi_def},
\begin{equation}
    \phi'(y) = \frac{\phi_1' \phi_2 - \phi_1 \phi_2'}{\phi_2^2} ,
    \label{phi_prime_def_pos_case}
\end{equation}
with $\phi_1'(y)\phi_2(y) = f'[u^{-1}(y)]$ and
$\phi_1(y)\phi_2'(y) = f[u^{-1}(y)]\, u''[u^{-1}(y)]/u'[u^{-1}(y)]$, so that
\begin{equation}
    \phi'(y) = \frac{1}{\{u'[u^{-1}(y)]\}^2} \left\{ f'[u^{-1}(y)] - f[u^{-1}(y)] \frac{u''[u^{-1}(y)]}{u'[u^{-1}(y)]} \right\} .
    \label{phi_prime_expr_pos_case}
\end{equation}
At $y=0$, since $u^{-1}(0)=t_0$,
\begin{equation}
    \phi'(0) = \frac{1}{[u'(t_0)]^3} \left[ u'(t_0) f'(t_0) - u''(t_0) f(t_0) \right] .
    \label{phi_prime_zero_pos_case}
\end{equation}
Substituting into~\eqref{I_deriv_phi_expr} and rewriting $-f'(t_0)$ and $f(t_0)$ as integrals
against $\delta'(t-t_0)$ and $\delta(t-t_0)$ gives the distributional identity
\begin{equation}
    \delta' [u(t)] = \frac{1}{\lvert u'(t_0) \rvert^3} \left[ u'(t_0)\, \delta'(t-t_0) + u''(t_0)\, \delta(t-t_0) \right] ,
    \label{delta_prime_u_t_proof_pos_case}
\end{equation}
using $u'(t_0)=\lvert u'(t_0)\rvert$ in this case.

\subsubsection{Case \texorpdfstring{$u'[u^{-1}(y)] < 0$}{2}}

Here $\phi_2(y) = -u'[u^{-1}(y)]$ and $u'(t_0)<0$. The computation is identical up to an
overall sign,
\begin{equation}
    \phi'(0) = - \frac{1}{[u'(t_0)]^3} \left[ u'(t_0) f'(t_0) - u''(t_0) f(t_0) \right] ,
    \label{phi_prime_zero_neg_case}
\end{equation}
which leads to
\begin{equation}
    \delta' [u(t)] = \frac{1}{\lvert u'(t_0) \rvert^3} \left[ u'(t_0)\, \delta'(t-t_0) + u''(t_0)\, \delta(t-t_0) \right] ,
    \label{delta_prime_u_t_proof_neg_case}
\end{equation}
using $-u'(t_0)=\lvert u'(t_0)\rvert$ in this case.

\subsubsection{Conclusion}

Equations~\eqref{delta_prime_u_t_proof_pos_case} and~\eqref{delta_prime_u_t_proof_neg_case}
coincide, so
\begin{equation}
    \delta' [u(t)] = \frac{1}{\lvert u'(t_0) \rvert^3} \left[ u'(t_0)\, \delta'(t-t_0) + u''(t_0)\, \delta(t-t_0) \right]
    \label{delta_prime_u_t_proof_gen}
\end{equation}
holds whenever $u'(t_0)\neq 0$. The extension to several simple zeros $t_i$ follows by
splitting $I$ into subintervals each containing a single zero.

\subsection{Proof of the expression of \texorpdfstring{$\delta''[u(t)]$}{the second derivative}}
\label{proof_sec_deriv_delta}

With the same setup and notation as in Sec.~\ref{proof_deriv_delta}, consider
\begin{equation}
    \mathcal{J} \equiv \int_I dt \, f(t)\, \delta'' [u(t)] = \int_J dy \, \phi(y)\, \delta'' (y) = \phi''(0) ,
    \label{J_sec_deriv_phi_expr}
\end{equation}
where $\phi(y)$ is defined in Eqs.~\eqref{phi_def}--\eqref{phi_1_phi_2_def}. We compute
$\phi''(y)$, again distinguishing the sign of $u'[u^{-1}(y)]$.

\subsubsection{Case \texorpdfstring{$u'[u^{-1}(y)] > 0$}{1}}

From Eq.~\eqref{phi_prime_expr_pos_case}, $\phi'(y) = \psi_1(y) - \psi_2(y)$ with
\begin{equation}
    \psi_1(y) \equiv \frac{f'[u^{-1}(y)]}{\{u'[u^{-1}(y)]\}^2} , \qquad \psi_2(y) \equiv \frac{f[u^{-1}(y)]\, u''[u^{-1}(y)]}{\{u'[u^{-1}(y)]\}^3} .
    \label{psi_1_2_def_pos_case}
\end{equation}
Differentiating,
\begin{equation}
    \psi_1'(y) = \frac{f''[u^{-1}(y)]\, u'[u^{-1}(y)] - 2 f'[u^{-1}(y)]\, u''[u^{-1}(y)]}{\{u'[u^{-1}(y)]\}^4} ,
    \label{psi_1_prime_expr_pos_case}
\end{equation}
\begin{align}
    \psi_2'(y) ={}& \frac{f'[u^{-1}(y)]\, u''[u^{-1}(y)] + f[u^{-1}(y)]\, u'''[u^{-1}(y)]}{\{u'[u^{-1}(y)]\}^4} \nonumber\\
    &- 3 \frac{f[u^{-1}(y)] \left\{ u''[u^{-1}(y)] \right\}^2}{\{u'[u^{-1}(y)]\}^5} ,
    \label{psi_2_prime_expr_pos_case}
\end{align}
so that
\begin{align}
    \phi''(y) ={}& \frac{ f''[u^{-1}(y)]\, u'[u^{-1}(y)] - 3 f'[u^{-1}(y)]\, u''[u^{-1}(y)] - f[u^{-1}(y)]\, u'''[u^{-1}(y)] }{\{u'[u^{-1}(y)]\}^4} \nonumber\\[0.2cm]
    &+ 3 \frac{f[u^{-1}(y)] \left\{ u''[u^{-1}(y)] \right\}^2}{\{u'[u^{-1}(y)]\}^5} .
    \label{phi_dbl_prime_expr_pos_case}
\end{align}
At $y=0$,
\begin{align}
    \phi''(0) = \frac{1}{[u'(t_0)]^5} \Big\{ &u'(t_0)\big[ u'(t_0) f''(t_0) - 3 u''(t_0) f'(t_0) - u'''(t_0) f(t_0) \big] \nonumber\\
    &+ 3 [u''(t_0)]^2 f(t_0) \Big\} .
    \label{phi_dbl_prime_zero_pos_case}
\end{align}
Substituting into~\eqref{J_sec_deriv_phi_expr} and re-expressing $f''(t_0)$, $-f'(t_0)$,
$f(t_0)$ as integrals against $\delta''$, $\delta'$, $\delta$ gives
\begin{equation}
    \delta'' [u(t)] = \frac{1}{\lvert u'(t_0) \rvert^5} \Big\{ u'(t_0)\big[ u'(t_0) \delta''(t-t_0) + 3 u''(t_0) \delta'(t-t_0) - u'''(t_0) \delta(t-t_0) \big] + 3 [u''(t_0)]^2 \delta(t-t_0) \Big\} ,
    \label{delta_dbl_prime_u_t_proof_pos_case}
\end{equation}
using $u'(t_0)=\lvert u'(t_0)\rvert$.

\subsubsection{Case \texorpdfstring{$u'[u^{-1}(y)] < 0$}{2}}

In this case $\phi'(y) = -[\psi_1(y)-\psi_2(y)]$, the negative-case analog of
Eq.~\eqref{phi_prime_expr_pos_case}, with the same $\psi_1,\psi_2$ as in
Eq.~\eqref{psi_1_2_def_pos_case}. The result of the previous
case applies up to the overall sign, and using $-u'(t_0)=\lvert u'(t_0)\rvert$ one obtains the
same identity,
\begin{equation}
    \delta'' [u(t)] = \frac{1}{\lvert u'(t_0) \rvert^5} \Big\{ u'(t_0)\big[ u'(t_0) \delta''(t-t_0) + 3 u''(t_0) \delta'(t-t_0) - u'''(t_0) \delta(t-t_0) \big] + 3 [u''(t_0)]^2 \delta(t-t_0) \Big\} .
    \label{delta_dbl_prime_u_t_proof_neg_case}
\end{equation}

\subsubsection{Conclusion}

Equations~\eqref{delta_dbl_prime_u_t_proof_pos_case}
and~\eqref{delta_dbl_prime_u_t_proof_neg_case} coincide, so the identity holds for
$u'(t_0)\neq 0$, and extends to several simple zeros as before.

\section{Late-time expansion for Model II}
\label{Appendix:LateTime}

We sketch the perturbative derivation of Eqs.~\eqref{eq:geff_ff} in the regime $t\gg\tau$ and
$\omega_0\tau\ll 1$. Writing the equation of motion as
\begin{equation}
\ddot x_c(t)+\frac{\omega_0^2\tau^2}{(t+\tau)^2}\,x_c(t)=g,
\qquad x_c(0)=\dot x_c(0)=0,
\label{eq:pert_eq}
\end{equation}
we set $\varepsilon:=(\omega_0\tau)^2$ and expand
$x_c(t)=x_0(t)+\varepsilon x_1(t)+O(\varepsilon^2)$ with $x_j(0)=\dot x_j(0)=0$. At zeroth
order $x_0(t)=\tfrac{g}{2}t^2$, $\dot x_0(t)=gt$. At first order,
\begin{equation}
\ddot x_1(t)=-\frac{x_0(t)}{(t+\tau)^2}=-\frac{g}{2}\frac{t^2}{(t+\tau)^2}.
\label{eq:x1eq}
\end{equation}
For $t\gg\tau$, $t^2/(t+\tau)^2 = 1+O(\tau/t)$, so
$\ddot x_1(t)=-\tfrac{g}{2}+O(g\tau/t)$, and integrating twice gives
$x_1(t)=-\tfrac{g}{4}t^2+O(g\tau t)$, $\dot x_1(t)=-\tfrac{g}{2}t+O(g\tau)$. Hence
\begin{align}
x_c(t) &= \frac{g}{2}t^2\left(1-\frac{1}{2}\omega_0^2\tau^2\right)
+O\!\left(g\,\omega_0^2\tau^3 t\right)+O\!\left((\omega_0\tau)^4 t^2\right),
\label{eq:xc_proved}\\
v_c(t) &= gt\left(1-\frac{1}{2}\omega_0^2\tau^2\right)
+O\!\left(g\,\omega_0^2\tau^3\right)+O\!\left((\omega_0\tau)^4 t\right),
\label{eq:vc_proved}
\end{align}
which is the $O(\tau^2)$ correction quoted in Eq.~\eqref{eq:geff_ff}.

For the width, with $\alpha:=\hbar/(2m\sigma^2)$,
$\sigma(t)=\sigma\sqrt{G_1(t)^2+\alpha^2 G_2(t)^2}$, where $G_1,G_2$ solve the homogeneous
form of Eq.~\eqref{eq:pert_eq}. Setting $u=t+\tau$, $k=\omega_0\tau$, $r=u/\tau=1+t/\tau$, the
exact solutions for $k\neq1/2$ are power laws $u^{m_\pm}$ with $m_\pm=(1\pm\Delta)/2$,
$\Delta=\sqrt{1-4k^2}$, giving
\begin{equation}
G_1(t)=\frac{1}{\Delta}\!\left[m_+\,r^{m_-}-m_-\,r^{m_+}\right],\qquad
G_2(t)=\frac{\tau}{\Delta}\!\left[r^{m_+}-r^{m_-}\right].
\label{eq:G1G2_exact_tau}
\end{equation}
Expanding for $k\ll1$,
\begin{align}
G_1(t) &= 1+k^2\!\left(\ln r+1-r\right)+O(k^4),
\label{eq:G1_expand}\\
G_2(t) &= t+\tau k^2\!\left[\,2(r-1)-(r+1)\ln r\,\right]+O(k^4),
\label{eq:G2_expand}
\end{align}
and substituting into $\sigma(t)$ yields
\begin{equation}
\sigma(t)=\sigma\,\sqrt{1+\alpha^2 t^2}\left[
1+\frac{k^2}{1+\alpha^2 t^2}
\left(
\ln r+1-r
+\alpha^2 t\,\tau\big\{2(r-1)-(r+1)\ln r\big\}
\right)
+O(k^4)
\right].
\label{eq:sigma_expand_general}
\end{equation}
In the ballistic regime $\alpha t\gg1$ this reduces to
\begin{equation}
\sigma(t)\simeq \frac{\hbar t}{2m\sigma}\left[
1+k^2\,\frac{\tau}{t}\big\{2(r-1)-(r+1)\ln r\big\}+O(k^4)
\right],
\label{eq:sigma_expand_ballistic}
\end{equation}
so the leading correction to the free-particle width is of relative order
$(\omega_0\tau)^2\ln(t/\tau)$ for $t\gg\tau$, provided $(\omega_0\tau)^2\ln(t/\tau)\ll1$, as stated below
Eq.~\eqref{eq:sigma_free_ff}.

\end{document}